 \documentclass[12pt,a4paper]{article}
 \usepackage{epsf}
\begin{document}
\title{The flavor-changing rare top decays $t\rightarrow c V V$ in
topcolor-assisted technicolor theory }
\author{Chongxing Yue$^{(a)}$, Gongru Lu$^{(a)}$,
 Qingjun Xu$^{b}$, Guoli Liu$^{b}$, Guangping Gao$^{b}$
 \\ {\small a: CCAST (World
 Laboratory) P.O. BOX 8730. B.J. 100080 P.R. China}\\
  \small{and College of Physics and Information Engineering,}\\
 \small{Henan Normal University,
 Xinxiang  453002. P.R.China} \\
 {\small b:College of Physics and Information Engineering,}\\
 \small{Henan Normal University,
 Xinxiang  453002. P.R.China}
\thanks{This work is supported by the National Natural Science
 Foundation of China, the Excellent Youth Foundation of Henan Scientific
 Committee; and Foundation of Henan Educational Committee.}
\thanks{E-mail:cxyue@pbulic.xxptt.ha.cn} }
\date{\today}
\maketitle
\begin{abstract}
\hspace{5mm}In the framework of topcolor-assisted technicolor
(TC2) theory, we calculate the contributions of the scalars( the
neutral top-pion $\pi_{t}^{0}$ and the top-Higgs $h_{t}^{0}$ ) to
the flavor-changing rare top decays $t\rightarrow c V V$($V= W,
g$, $\gamma$ or $Z$). Our results show that $h_{t}^{0}$ can
enhance the standard model $B_{r}^{SM}(t\longrightarrow cWW)$ by
several orders of magnitude for most of the parameter space. The
peak of the branching ratio resonance emerges when the top-Higgs
mass is between $2m_{W}$ and $m_{t}$. The branching ratio $
B_{r}(t\rightarrow c W W )$ can reach $ 10^{-3}$ in the narrow
range.
\end {abstract}

\newpage
    The large value of the top quark mass offers the possibility
that it plays a special role in current particle physics. Indeed,
the properties of the top quark could reveal information on flavor
physics, electroweak symmetry breaking(EWSB) as well as new
physics beyond the standard model(SM)\cite{x1}. One of these
consequences is that the flavor-changing rare top decays which are
very small due to the GIM-suppressed in the SM can be used to
detect new physics. This fact has lead to a lot of theoretical
activity involving the rare top decays within some specific models
beyond the SM\cite{x2}.

    The strong top dynamical symmetry breaking models, such as
topcolor-assisted technicolor(TC2) models\cite{x3} and top see-saw
models\cite{x4}, are attractive because they explain the large top
quark mass and provide possible dynamical mechanism for breaking
electroweak symmetry. Such type of models generally predict light
composite scalars with large Yukawa couplings to the third
generation. This induces distinct new flavor mixing phenomena
which may be tested at both low and high energies \cite{x5,x6}.
For example, TC2 theory \cite{x3} predicts the existence of the
top-pions$ ( \pi _{t}^{\pm},\pi _{t}^{0}) $ and the neutral
CP-even state, called top-Higgs $ h_{t}^{0}$. These new particles
are most directly related to the dynamical symmetry breaking
mechanism. Thus, studying the possible signatures of these new
particles at high energy colliders will be of special interest.

  Ref[5] has pointed out that the Yukawa couplings of the scalars
 to charm and
bottom quarks can be large due to a significant mixing of the top
and charm quarks. Furthermore, the neutral scalars($\pi_{t}^{0}$
and $h_{t}^{0}$) can couple to a pair of gauge bosons through the
top quark triangle loop in an isospion violating way\cite{x7}.
The main difference between the neutral top-pion $\pi _{t}^{0}$
 and top-Higgs $h_{t}^{0}$ is that $h_{t}^{0}$ can couple to gauge boson
 pairs WW and ZZ at tree level, which is similarly to that of the SM
 Higgs H. Thus, the neutral scalars may have significant
 contributions to the rare top decays $t\rightarrow c V V(V=W,g,
\gamma$ or $Z$)[$t\rightarrow c W W$ (only for $h_{t}^{0}$),
$t\rightarrow c g g$, $t\rightarrow c\gamma\gamma$ and
$t\rightarrow c Z \gamma$].

     The top quark mass has been measured \cite{x8} by
reconstructing the decay products of top pairs produced at the
Tevatron. The combined measurement from CDF and D0 is $m_{t}\doteq
173.4\pm 5.1 $GeV. This implies that the rare top decay
$t\longrightarrow cWW$ is allowed. However, this process is
occurring near threshold and is highly phase space suppressed.
Within the SM, the decay channel $t\longrightarrow cWW $ is also
highly GIM-suppressed $(B_{r}^{SM}(t\longrightarrow c W W)\approx
10^{-13}\cite{x9})$, which can not be observed in the future high
energy colliders. So, studying such rare top decay will be very
useful to detect the effects of new physics.

   In this letter, we first calculate the contributions of the top-Higgs
   $h_{t}^{0}$ to the rare decay channel $t\longrightarrow cWW$. We find
   that the peak
 of the branching ratio resonance emerges when the top-Higgs mass is
 between $2m_{W}$ and $m_{t}$. For  $m_{h^{0}_{t}}$ = 165 GeV, the
 value of $B_{r}(t\longrightarrow c W W)$ is $3\times 10^{-3}$ for
 $\varepsilon$ = 0.01 and $5.6\times 10^{-2}$ for $\varepsilon$ = 0.08.
 We further estimate the partial widths of the rare top decays
$t\rightarrow c g g$, $t\rightarrow c\gamma\gamma$ and
$t\rightarrow c Z \gamma$ contributed by the neutral scalars (
$\pi _{t}^{0}$ and $h_{t}^{0}$). The new contributions can enhance
the SM partial widths by several orders of magnitude. Even so, it
is very difficult to detect the possible signatures of the neutral
scalars via these flavor-changing processes.

   To solve the phenomenological difficulties of traditional TC
 theory, TC2 theory \cite{x3} was proposed by combing TC
 interactions with the topcolor interactions for the third
  generation at the scale of about 1 TeV. In TC2 theory,
  the TC interactions play a main role in breaking the
 electroweak symmetry. The ETC interactions give rise to the
masses of the ordinary fermions including a very small portion of
the top quark mass, namely $\varepsilon m_{t}$ with a model
dependent parameter $\varepsilon \ll 1$. The topcolor interactions
also make small contributions to the EWSB, and give rise to the
main part of the top quark mass, $(1-\varepsilon)m_{t}$. So, for
TC2 theory, there is the following relation:
\begin{equation}
  \nu_{\pi}^{2}+F_{t}^{2}=\nu_{w}^{2},
\end{equation}
 where $\nu_{\pi}$ represents the contributions of the TC or other
 interactions to the EWSB, $F_{t} \approx 50GeV $ is decay constant of
 the scalars predicted by TC2 theory, and $\nu _{w}$ = $\nu
 / \sqrt{2} \approx 174GeV$. Thus, the majority of the masses of
  gauge bosons W and Z come from the technifermion condensate.

    For TC2 models, the underlying interactions, i.e. topcolor
 interactions, are non-universal and therefore do not possess a
 GIM mechanism. When the non-universal
interactions are written in the mass eigen-states, it may lead to
the flavor changing coupling vertices of the new gauge bosons,
such as $Z't c, Z'\mu e, Z'\mu\tau$. Thus, the new gauge boson
$Z'$ have significant contributions to some lepton flavor changing
processes\cite{x10}. Furthermore, the neutral scalars predicted by
this kind of models have the flavor changing scalar coupling
vertices. The coupling of the neutral scalars $ S $ ( $
\pi_{t}^{0}$ or $ h_{t}^{0}$) to the ordinary fermions can be
written as\cite{x3,x5}:
\begin{equation}  S t\overline{t}: \frac{im_{t}}{\sqrt{2}F_{t}}
\frac{\sqrt{\nu_{w}^{2}-F_{t}^{2}}}{\nu_{w}} K_{UR}^{tt},
\hspace{5mm}
 S \overline{t}c:  \frac{im_{t}}{\sqrt{2}F_{t}}
 \frac{\sqrt{\nu_{w}^{2}-F_{t}^{2}}}{\nu_{w}}K_{UR}^{tc}.
 \end{equation}
 Ref.\cite{x5} has shown that the value of $K_{UR}^{ij}$ can be
 taken as:
 \begin{equation}
  K_{UR}^{tt} = 1-\varepsilon,\hspace{5mm} K_{UR}^{tc} \leq
 \sqrt{2\varepsilon-\varepsilon^{2}}.
 \end{equation}
The couplings of the scalars to the bottom quark can be
approximately written as :
\begin{equation} S b\overline{b} : \frac{i(m_{b}-m_{b}^{'})}{\sqrt{2}F_{t}}
\frac{\sqrt{\nu_{w}^{2}-F_{t}^{2}}}{\nu_{w}},
\end{equation}
where $m_{b}^{'}$ is the ETC generated part of the bottom-quark
mass. According to the idea of TC2 theory, the masses of the first
and second generation fermions are also generated by ETC
interactions. We have $\varepsilon
m_{t}=\frac{m_{c}}{m_{s}}m_{b}^{'}$ \cite{x11}. If we take
$m_{s}$=0.12GeV and $m_{c}$=1.2GeV, then we have
$m_{b}^{'}=0.1\times\varepsilon m_{t}$.

   The couplings of the neutral scalars to gauge boson pairs $ g g $,
$ \gamma\gamma $ or $ Z \gamma $ via the top quark triangle loop
are isospin violating. The general form of the effective $ S
-V_{1}-V_{2}$ couplings can be written as \cite{x7,x12}:
\begin{equation}
\frac{1}{1+\delta_{V_{1}V_{2}}}\frac{\alpha S_{S V_1V_2}}{\pi F_t}
S \epsilon_{\mu\nu\alpha\beta}(\partial^{\mu}
V_{1}^{\nu})(\partial^{\alpha}V_{2}^{\beta}),
\end{equation}
where $V_{1}^{\nu}$ and $V_{2}^{\beta}$ represent the field
operators of the gauge bosons. The anomalous factors $S_{S
V_{1}V_{2}}$ are model dependent. They have been given in
Refs.[6,12].

    The neutral top-pion $\pi_{t}^{0}$ can not couple to gauge boson pairs
$ W W $ and $ Z Z $ at tree level. The couplings of the top-Higgs
$h_{t}^{0}$ to the electroweak gauge bosons at tree level are
suppressed by the factor $F_{t}/\nu_{w}$ with respect to that of
the SM Higgs. For the top-Higgs $h_{t}^{0}$, we have
 \begin{equation}
  h_{t}^{0}WW : \frac{i F_{t}}{\nu_{w}} gm_{W}g_{\mu
         \nu} ,\hspace{5mm}
 h_{t}^{0}ZZ : \frac{i F_{t}}{\nu_{w}}\frac{gm_{Z}}{cos\theta_{W}}
 g_{\mu \nu}.
 \end{equation}

    From above discussion,  we can see that the top-Higgs
$h_{t}^{0}$ may have significantly contributions to the rare top
 quark decay channel $t\longrightarrow cWW$. The relative
 amplitude is:
\begin{eqnarray}
M(t\longrightarrow c W W) &= &  \frac{m_{t}}{\sqrt{2}F_{t}}
\frac{\sqrt{\nu_{w}^{2}-F_{t}^{2}}}{\nu_{w}} K_{UR}^{tc} \frac
{F_{t} g m_{W}}{\nu_{w}} \overline{u}(p_{c})\gamma_{5}u(p_t)
\nonumber \\
 &&\frac{1}{K^{2}-m_{h^{0}_{t}}^{2} + im_{h^{0}_{t}} \Gamma_{total}}
 \varepsilon_{\mu}(k_{1},\lambda_{1})g^{\mu\nu}\varepsilon_{\nu}
 (k_{2},\lambda_{2}),
\end{eqnarray}
 where the four momenta $K$ is given by
\begin{equation}
K=P_{t}-P_{c}=k_{1}+k_{2}.
\end{equation}
Where $k_{i} $ is the four momenta of the gauge boson $ W $.
 For $150GeV\leq m_{h^{0}_{t}}\leq 350GeV$, the total
decay width of the top-Higgs $h_{t}^{0}$ can be written as :
\begin{eqnarray}
 \Gamma_{total}&=&\Gamma(h_{t}^{0}\longrightarrow b\overline{b})
      +\Gamma(h_{t}^{0}\longrightarrow gg)
        +\Gamma(h_{t}^{0}\longrightarrow \gamma\gamma) \nonumber \\
   &+&\Gamma(h_{t}^{0}\longrightarrow Z\gamma)
       +\Gamma(h_{t}^{0}\longrightarrow \overline{t}c)
   (for \hspace{5mm}m_{h_{t}^{0}}\geq m_{t}+m_{c}) \nonumber \\
     &+&\Gamma(h_{t}^{0}\longrightarrow WW)
   (for \hspace{5mm} m_{h_{t}^{0}}\geq 2m_{W})\nonumber \\
    &+&\Gamma(h_{t}^{0}\longrightarrow ZZ)
  (for \hspace{5mm} m_{h_{t}^{0}}\geq 2m_{Z}).
 \end{eqnarray}
      The branching ratio $B_{r}(t\longrightarrow cWW)$ contributed
 by the top-Higgs $h_{t}^{0}$ is plotted in Fig.1 as a
 function of the top-Higgs mass $m_{h_{t}^{0}}$ for three values of
  the parameter $\varepsilon$. In Fig.1 we have assumed that the
  total top width is dominated by the decay channel $ t\longrightarrow
  W b $ and taken $ \Gamma(t \longrightarrow W b)= 1.56 GeV $\cite{x1}.
  The three-body phase space integral was performed numerically for
the parameter values $m_{W}=80.4GeV$, $m_{t}=175GeV$,
$m_{c}=1.2GeV$, $\alpha_{e}=\frac{1}{128}$, $\alpha_{s}=0.118$ and
$\sin\theta_{w}=0.2312$ \cite{x13}. From Fig.1 we can see that the
peak of the branching ratio $B_{r}(t\longrightarrow c W W)$
resonance emerges when $m_{h_{t}^{0}}$ is between $2m_{W}$ and
$m_{t}$. This is consisted with the results obtained in Ref.[14].
For  $m_{h_{t}^{0}}$=165GeV, the value of the
$B_{r}(t\longrightarrow c W W)$ is $3\times10^{-3}$ for
$\varepsilon$=0.01 and $5.6\times10^{-2}$ for $\varepsilon$=0.08.
The  $B_{r}(t\longrightarrow c W W)$ decreases rapidly in the
regions $ m_{h_{t}^{0}}<2m_{W}$ or $m_{h_{t}^{0}}>m_{t}$. However,
for most of the parameter space of the TC2 theory, the branching
ratio is several orders of magnitude larger than the
$B_{r}^{SM}(t\longrightarrow cWW)$.

    The amplitudes of the rare top decays $t\rightarrow c g g$,
 $t\rightarrow c\gamma\gamma$ and $t\rightarrow c Z \gamma$
 generated by the neutral top-pion $\pi_{t}^{0}$ can be written as:
\begin{eqnarray}
M(t\rightarrow c V V)&=&\frac{m_{t}}{\sqrt{2}F_{t}}
\frac{\sqrt{\nu_{w}^{2}-F_{t}^{2}}}{\nu_{w}}
K_{UR}^{tc}\frac{\alpha S_{\pi_{t}^{0}V_{1} V_{2}}}{2\pi F_{t}}
\\ \nonumber
&&\overline{u}(P_{c})\gamma^{5}u(P_{t})\frac{1}{K^{2}-m_{\pi_{t}}^{2}+i
m_{\pi_{t}}\Gamma}
(P_{V_{1}\mu}\epsilon_{V_{1}\nu}-P_{V_{1}\nu}\epsilon_{V_{1}\mu})
(P_{V_{2}}^{\mu}\epsilon_{V_{2}}^{\nu}
-P_{V_{2}}^{\nu}\epsilon_{V_{2}}^{\mu}),
\end{eqnarray}
where $\Gamma$ is the total decay widths of the neutral top-pion
$\pi_{t}^{0}$.

    The partial decay widths of the rare top decay channels
$t\rightarrow c V V$ are plotted in Fig.2 as functions of the
top-pion mass $m_{\pi_{t}}$ for $\epsilon=0.01$. In Fig.2, we have
taken the cut that the angle between photons or gluons is larger
than $15^{\circ}$ and the energy of photons or gluons
$E_{\gamma(g)}\geq20GeV$. From Fig.2 we can see that the partial
widths decrease as $m_{\pi_{t}}$ increasing in most of the
parameter space. If we assume that the part of the top quark mass
generated by the topcolor interactions makes up 99\% of $m_{t}$,
then we have $\Gamma(t\rightarrow c g g)\sim10^{-9}GeV$,
$\Gamma(t\rightarrow c\gamma\gamma)\sim10^{-10}GeV$ and
$\Gamma(t\rightarrow c Z\gamma)\sim10^{-10}GeV$.

     Ref.[15] has discussed the rare top decay channel $t\rightarrow c
H$ in the SM. Their results show that $B_{r}(t\rightarrow c
H)\approx9\times10^{-14}$ for $m_{H} = 100GeV$. The dominant decay
modes of the SM Higgs boson are $b\bar{b}$, $\tau\bar{\tau}$ and
$c\bar{c}$. The branching ratios $B_{r}^{SM}(H \rightarrow V V)$
are very small: $B_{r}^{SM}(H \rightarrow g g )\approx
5\times10^{-2}$, $ Br^{SM}(H \rightarrow \gamma \gamma )\sim
10^{-3}$ and $ B_{r}^{SM}(H \rightarrow Z \gamma) \sim
10^{-4}$\cite{x16}. Thus, the $B_{r}(t\rightarrow c V V)$
contributed by the neutral top-pion $\pi_{t}^{0}$ is larger than
that of the SM by several orders of magnitude.

   The contributions of $h_{t}^{0}$ to the rare top decays
$t\rightarrow c V V$( $V= g $, $\gamma $ or $ Z $ ) are similar to
that of $\pi_{t}^{0}$. Certainly, $h_{t}^{0}$ can couple to gauge
boson pair $W W$ and can give contributions to the rare top decays
$t\rightarrow c \gamma\gamma$ and $t\rightarrow c Z \gamma$ via
$W$ loops. This may enhance the branching ratios
$B_{r}(t\rightarrow c \gamma\gamma)$ and $B_{r}(t\rightarrow c Z
\gamma)$ relative to that of $\pi_{t}^{0}$. However, the coupling
$h_{t}^{0}W W$ is suppressed with respect to the case of the SM
Higgs boson $H$ by a factor $\frac{F_{t}}{\nu_{w}}$. Thus the
enhancement is very small. We can neglect the contributions of $W$
loops to the rare top decays $t\rightarrow c \gamma\gamma$ and
$t\rightarrow c Z \gamma$. The decay widths of the decays
$t\rightarrow c V V$ ( $V= g $, $\gamma $ or $ Z $ ) given by the
top-Higgs $ h^{0}_{t} $ approximately equal to that of the neutral
top-pion $\pi^{0}_{t} $.

    To assess the discovery reach of the rare top quark decays
in the future high energy colliders, Ref.[17] has roughly
estimated the following sensitivities  for $100fb^{-1}$ of
integrated luminosity :
\begin{equation}
LHC   :B_{r}(t\longrightarrow cX) \geq 5\times10^{-5} \nonumber
\end{equation}
\begin{equation}
 LC    :B_{r}(t\longrightarrow cX) \geq 5\times10^{-4} \nonumber
 \end{equation}
\begin{equation}
TEV33 :B_{r}(t\longrightarrow cX) \geq 5\times10^{-3}
\end{equation}
 Thus, the effects of $h_{t}^{0}$ on the rare top decay $t\longrightarrow
 c W W$ can be detected in the future high energy colliders. If
 it is not this case, we can conclude that the mass of the
 top-Higgs $h_{t}^{0}$ must be larger than 180GeV.

   The scalars predicted by the TC2 theory have large Yukawa couplings
 to the third family fermions and induce the new flavor changing
scalar couplings including the $t-c$ transitions for the neutral
scalars. Thus, the neutral scalars have significant contributions
to the rare top decay channels $t\rightarrow c V V$. If the mass
of the top-Higgs lies in the narrow range  $160GeV \leq
m_{h^{0}_{t}}\leq 180 GeV$, the rare top decay $t\rightarrow c W
W$ may be used to detect the signatures of the top-Higgs $
h_{t}^{0} $. For the neutral top-pion $ \pi_{t}^{0} $, we have to
use other processes to detect its possible signatures.
\newpage
\vskip 2.0cm
\begin{center}
{\bf Figure captions}
\end{center}
\begin{description}
\item[Fig.1:]The branching ratio $B_{r}(t\rightarrow c W W)$ as a
function of the top-Higgs mass $m_{h_{t}}$ for the parameter
$\epsilon= 0.01$ (solid line), $0.05$(dotted line) and $ 0.08
$(dashed line).
\item[Fig.2:]The partial decay widths $\Gamma(t\rightarrow c V V)$
 versus the mass $m_{\pi_{t}}$ for $\epsilon=0.01$.

\end{description}

\newpage
\begin{figure}
\begin{center}
\begin{picture}(300,200)(0,0)
\put(-50,0){\epsfxsize120mm\epsfbox{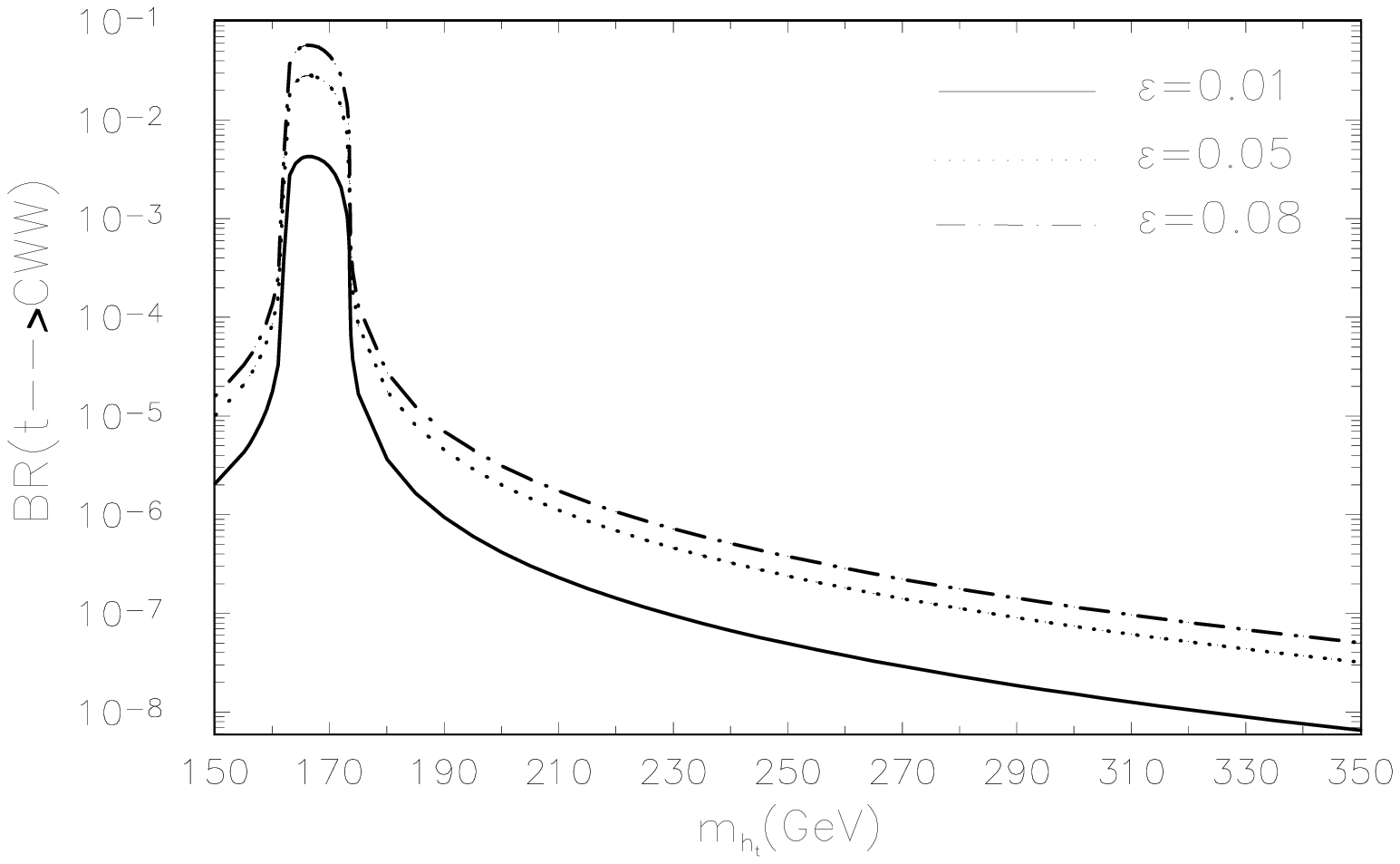}} \put(120,-10){Fig.1}
\end{picture}
\end{center}
\end{figure}
\begin{figure}[hb]
\begin{center}
\begin{picture}(300,200)(0,0)
\put(-50,0){\epsfxsize120mm\epsfbox{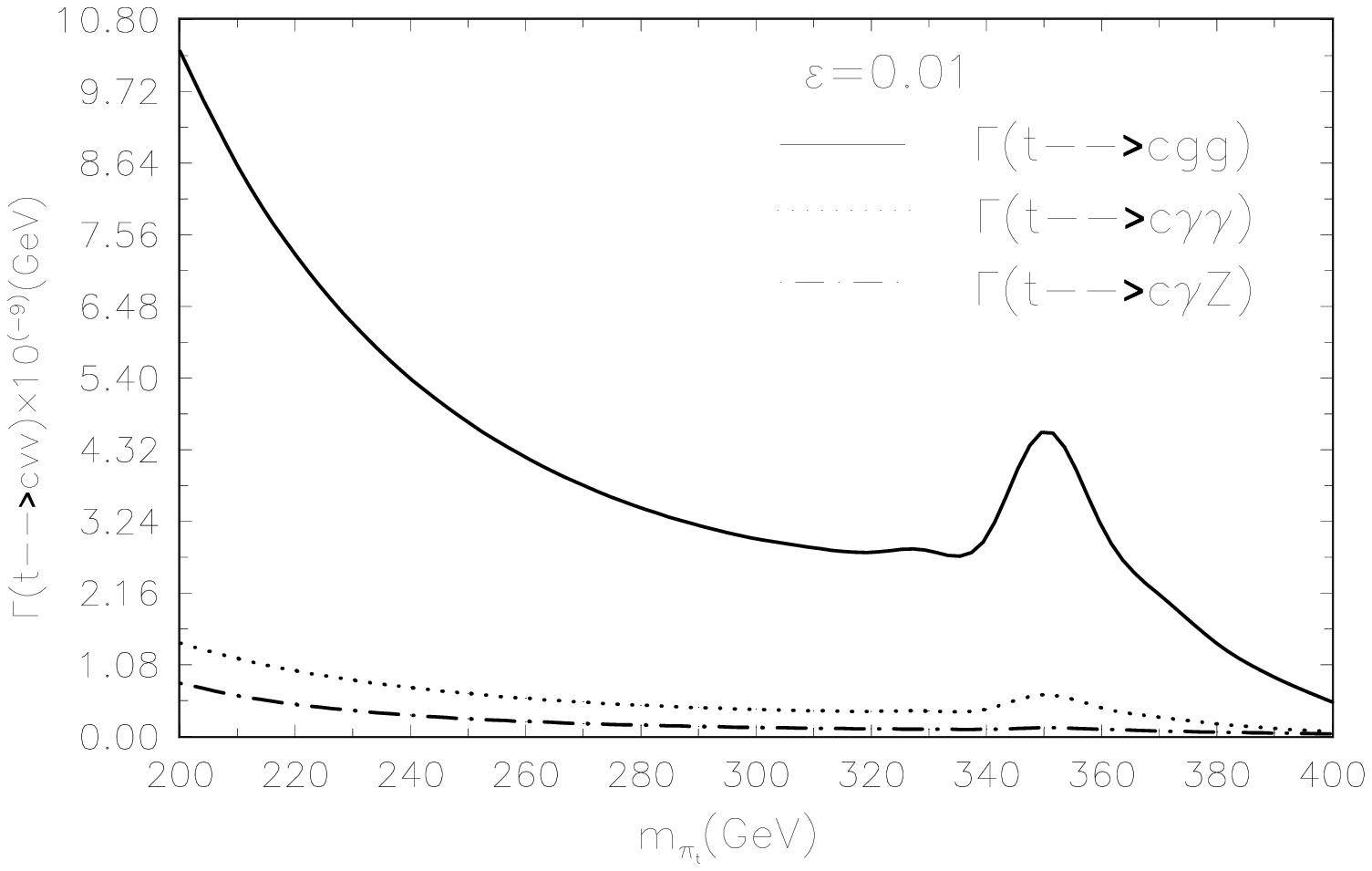}} \put(120,-10){Fig.2}
\end{picture}
\end{center}
\end{figure}

\end{document}